%% file: ms.tex
\documentclass[preprint,12pt]{aastex}

\shorttitle{Colliding Clouds: The BD+40\arcdeg~4124 Cluster}
\shortauthors{Looney et al.}

\newcommand{\bdstar}{$\rm BD+40\arcdeg~4124\ $}
\newcommand{\msun}{M$_{\sun}$}

\newcommand{\Jequals}[2]{\mbox{$J = {#1}\rightarrow{#2}$}}
\newcommand{\uv}{{\it u,v}}
\newcommand{\hii}{\ion{H}{2}\ }
\begin{document}

%\title{Clouds Colliding into the Cluster Environment Around BD+40\arcdeg~4124
%and a New Deeply Embedded Protostar}
\title{Colliding Clouds: The Star Formation
Trigger of the Stellar Cluster Around BD+40\arcdeg~4124}
\author{Leslie W. Looney\altaffilmark{1}, Shiya Wang\altaffilmark{1}, 
Murad Hamidouche\altaffilmark{1}, 
Pedro N. Safier\altaffilmark{2,3},\\ 
and Randolf Klein\altaffilmark{4,5}}

\altaffiltext{1}{Department of Astronomy, 
University of Illinois, 1002 W. Green St., Urbana, IL 61801}
\altaffiltext{2}{Department of Astronomy,
University of Maryland, College Park, MD 20742}
\altaffiltext{3}{Current Address: S \& J Solutions LLC}
\altaffiltext{4}{Max-Planck-Institut f\"ur 
extraterrestrische Physik, Garching, 85741, Germany}
\altaffiltext{5}{Current Address: Department of Physics, 
University of California, Berkeley, CA}
%\altaffiltext{6}{Department of Astronomy, California Institute of 
%Technology, Pasadena, CA 91125}

\begin{abstract}
We present BIMA and SCUBA observations of the young
cluster associated with \bdstar in the dense molecular gas
tracer CS~J~=~2~$\rightarrow$~1 and
the continuum dust continuum at 
$\lambda$~=~3.1~mm and  $\lambda$~=~850~$\mu$m.
The dense gas and dust in the system
is aligned into a long ridge morphology extending $\sim$0.4 pc with
16 gas clumps of estimated masses
ranging from 0.14 - 1.8 M$_\odot$.
A north-south variation in the CS center line velocity 
can be explained with a two cloud model.
We posit that the \bdstar stellar cluster formed from a 
cloud-cloud collision.
The largest linewidths occur near V1318~Cyg-S, a
massive star affecting its natal environment.
In contrast, the dense gas near the other, more evolved, massive stars
displays no evidence for disruption;
the material must either
be processed into the star, dissipate, or relax, fairly quickly.
The more evolved low-mass protostars are more likely to be found near
the massive stars.
If the majority of low-mass stars are coeval,
the seemingly evolved low-mass protostars
are not older: the massive stars have eroded their structures.
Finally, at the highest resolution,
the $\lambda$ = 3.1 mm dust emission is resolved into
a flattened structure 3100 by 1500 AU with an estimated mass of 3.4 M$_\odot$.
The continuum and CS emission are
offset by 1$\farcs$1 from the southern binary source.
A simple estimate of the extinction due to the
continuum emission structure is A$_V$ $\sim$ 700 mag.
From the offset and as the southern source is detected in the
optical, the continuum emission is
from a previously unknown very young, intermediate-mass,
embedded stellar object.
\end{abstract}

\keywords{ISM: clouds---ISM: jets and outflows---radio lines:
  ISM---stars: formation---stars:
  individual (\bdstar,  V1318~Cyg, V1318~Cyg-S, V1318~Cyg-N)}

\section{Introduction}

Over the last few decades, a good broad-brush picture of isolated, 
low-mass,
Sun-like star formation \citep[e.g.,][]{shuppiii} has been developed.
However, these theories are unlikely to explain properly the formation of
our Sun or indeed the majority of other stars in the Galaxy. For our Sun
in particular, the isotopic evidence strongly suggests that it formed
in a location that was enriched either by a stellar wind or a recent
supernova \citep[e.g.,][]{lee1977}, hence in or near a cluster environment.
New results show that the early solar system contained significant amounts
of the short-lived $^{60}$Fe radionuclide \citep[e.g.,][]{tachibana2003},
probably from a nearby supernova \citep[$\sim$ 1.5 pc distance,][]{supernova}. 

The suggestion that our Sun might have been born in a cluster
environment, should not be too surprising, as there is good
evidence that the majority of stars in the Galaxy formed in clusters
\citep[e.g.,][]{carpenter2000,ladalada2003}.  Giant molecular clouds
in the Galaxy provide the ideal environment for star formation,
comprising dense molecular gas and dust that provide the raw material
\citep[e.g.][]{Zuckerman74}.  But how do we expect low-mass star
formation to proceed in a cluster?  According to the standard model
\citep[e.g.][]{shuadaliz87}, it
is a relatively long process requiring a core to exceed its Jeans
mass, initiate self-similar collapse, followed by multiple fragmentation
down to star formation with the Initial Mass Function (IMF) 
stellar distribution.  In this model,
higher mass stars form quickly and low mass stars form slowly.

However, in clusters the environment is much more complicated.
Clusters are often associated with one or many massive stars accompanied
by a large number of lower mass stars.  In fact, the typical star
forming system is actually a low stellar density cluster, including young
intermediate to high mass star or stars (3-20 $M_{\odot}$) surrounded by
ten to a few hundred of low-mass protostars. This trend for massive stars
to be more and more gregarious, is first associated with stars at the
Herbig Ae/Be stage \citep[e.g.][]{lynne95,palla95}.  In such an environment,
the standard model has to be modified for the large influence that
the massive star will have on low-mass star formation and evolution.
The low-mass star may evolve faster than in isolation.
It has been posited that stellar formation in a large cluster, with a
combined mass of a few hundred $M_\odot$, evolves faster by minimizing
the effects of magnetic fields \citep{shuadaliz87}.  Are low-mass stars
formed coevally with massive stars during a trigger event?  Are low-mass
stars triggered by the high mass star?  
%Or, more likely some combination
%of the two?
%This implies that the standard
%model of star formation probably fails for most stars.  We need to foster
%observations and models that address the competition for material,
%the increased turbulent motion, and the stellar interactions through
%gravity, winds, and photon heating/ionization that are expected in
%cluster environments.

In order to better understand the differences between isolated low-mass
star formation and low-mass star formation in a cluster with massive
stars, it is important to understand the most simple case of small
clusters, or groups, around intermediate-mass stars.
One of the main points is to develop a better picture of how
the most massive members of the cluster influence the formation of the
lower-mass members by probing the dense molecular gas, dust, and stellar
population.
An excellent
example is the small cluster surrounding the intermediate-mass/massive
($\sim$ 9 $M_\odot$), young \citep[$\sim$ 1 Myr;][]{lynne95} 
star \bdstar, one of the original Ae/Be stars studied by
\cite{herbig}.  Although the main star has a spectral type of B2 Ve 
and is hot enough ($T_{\rm eff} > 22\,000$~K), 
there is no observational evidence for an \hii region \citep{skinner93}.
Located in the Cygnus arm at a distance of about one kiloparsec,
\bdstar is the optically brightest member of a small group of young stars
that include V1686~Cyg and the binary V1318~Cyg, all of which have
infrared excesses \citep{strom72} and are very young \citep{cohen72}.

\bdstar defines the center of a partially embedded aggregate of at
least 33 stars, 80\% of which are located within a projected distance
of $0.15{\rm\,pc}$ from the central star \citep{lynne95}.
Single-dish maps of CO, $^{13}$CO, C$^{18}$O and CS,
\citep{lynne95,palla95} and NH$_3$ \citep{fuente90},
show that this relatively isolated and dense
($n \ga 1000\,{\rm stars\,pc^{-3}}$) cluster is still associated with more
than $\sim 300\,M_\odot$ of clumpy, dense molecular
material. Therefore, ongoing star-formation is still possible in
this region.  In fact, \cite{lynne95} concluded that
high- and  low-mass stars are still forming,
with the balance in favor of high-mass stars.

This small and young group of stars is interesting for
two main reasons.
First, the parent molecular core is isolated from any large
star-forming complex, and, therefore, this region is an ideal site
to investigate the conditions in which a single high-mass star,
capable of becoming a supernova
($M_\ast \ga 8\,$\msun) forms, 
and its role, if any, in the formation of its lower-mass
siblings. Second, \cite{lynne95} concluded that star
formation in this region was triggered by an external event,
based on morphological arguments. However,
no evidence can be found for any of the standard mechanisms for
induced star formation (e.g. nearby expanding \hii
regions, supernova remnants, or nearby OB associations).
%even though several clouds with \vlsr = 2, 6, 8, 12.5, and
%14\,\kms\ are detected in \twelveco\ in this region of the sky
%\citep{lor77,canto84} --with the \vlsr = 8\,\kms\ component
%being typically attributed to the \bdstar association --the lack of
%emission at intermediate velocities argues against a cloud-collision
%scenario. Therefore, by studying this region one may hope to find
%leftover signatures from the original triggering event.

To unravel the star-formation history
of this region it is first necessary to
identify the sites of ongoing-- and potentially future-- star
formation and to study the kinematic signatures and
physical conditions of these sites. 
%This requires
%high  spatial- and spectral-resolution imaging of the dense material
%using a high density tracer like CS. 
The observations focused on the CS~($2\rightarrow 1$) transition,
as a tracer of dense gas.
Due to its high critical density, 
\citep[$n_{H_2}\sim10^{5}\rm\, cm^{-3}$, e.g.][]{irvine87},
CS~($2\rightarrow 1$) is thermalized
in the dense gas of star-forming cores.  In other words,
CS emission is a good method for revealing dense, star forming clumps.
To date, the available
observations at millimeter
wavelengths are of moderate spatial resolution-- they were all obtained with
single-dish telescopes. 
In this paper, we present high spatial- and spectral-resolution
data obtained with
the BIMA array in the $J = 2\rightarrow 1$ transition of CS and 
dust continuum emission that probe the
dense gas and dust in the young cluster on spatial scales of 1000's 
to 10,000's of AU.

\section{Observations}

\bdstar was observed in three configurations (B, C, and D) of the
10-element BIMA Array\footnotemark\ \citep{bima}.
The uncertainty in the absolute amplitude calibration is
estimated to be 20\%;
all BIMA flux uncertainty discussed is
the statistical uncertainty and does not account for
any amplitude calibration error.
The CS \Jequals{2}{1} ($\nu$ = 97.981 GHz)
observations in B and D configurations were
acquired in 2003 March and August.  The digital correlator was configured with 
the line window having a velocity range of 69 km/s with 
0.27 km/s channels and two
600~MHz bands for continuum.
The system temperatures during the observations ranged from
170-600 K (Single Side-Band, SSB).
In addition, we also used archival data from 
1996 August of the 9-element array in C configuration.
The digital correlator for the archival data was configured
similarly but with only 500~MHz bands for continuum.
The system temperatures for the archival data ranged
from  310-500 K (SSB).

%The C$^{18}$O \Jequals{1}{0} and $^{13}$CO \Jequals{1}{0} transitions
%were observed simultaneously in another observing campaign.
%The three configurations were observed in February, April, and June 2004.
%The digital correlator was configured with two line windows for the two transitions,
%each having a velocity range of 138 km/s at 1 km/s channels and
%two 100~MHz bands for continuum.
%The system temperatures during the observations ranged from
%290-600 K (SSB).
%
The data were reduced with the MIRIAD package
\citep{miriad}.
The observations span \uv\ distances from 2 k$\lambda$
to 75 k$\lambda$, providing brightness distribution information 
on spatial scales from 40$\arcsec$ to 1$\farcs$1.
In order to display this information in the image plane, we
mapped the data using all three configurations with two different \uv\
weighting schemes that stress structures
on spatial scales of roughly 6$\arcsec$ and 2$\arcsec$.
These resolutions were obtained
with robust weighting \citep{robust} values of 0.2 and
-0.5, respectively.

\bdstar continuum submillimeter observations were obtained from the
public archive data of the Submillimetre Common User Bolometer Array
(SCUBA) instrument on the James Clerk Maxwell Telescope (JCMT). The
observations were carried-out using the long wavelength array at 850
$\mu m$ and the short wavelength array at 450 $\mu m$. The source
was observed in 1997 September. 
The observing conditions were poor; the average sky opacities
during the observation were $\tau_{850\mu m}$ = 0.91 at 850 $\mu m$
and $\tau_{450\mu m}$ = 5.19 at 450 $\mu m$.
The data were reduced
using the SCUBA User Reduction Facility procedures \citep{surf}.
Specific care was taken to remove
sky noise and saturated bolometers. The planet Uranus and CRL
2688 were used as flux calibrators. 
The data were reduced for both long
and short wavelengths to ensure consistency, although 
only the long wavelength array data are used.
An uncertainty in the absolute flux calibration as 
20\% is estimated; all SCUBA flux uncertainty discussed is
the statistical uncertainty and does not account for
any amplitude calibration error.

\footnotetext{ The BIMA Array was operated by the Berkeley Illinois Maryland
Association under funding from the National Science Foundation.
BIMA has since combined with the Owens Valley Radio Observatory 
millimeter interferometer, moved to a new higher site, 
and is being recommissioned as the Combined
Array for Research in Millimeter Astronomy (CARMA) in early 2006.
}

\section{Results}

The BIMA data provide the highest resolution observations of
dense gas in the \bdstar cluster to date.
The data were obtained in three configurations of the
interferometer, each probing structure on varying size-scales.
The data was combined and weighted to emphasize
structures on two different size scales-- $\sim$6$\arcsec$ and 2$\arcsec$, or
equivalent to a spatial resolution of 6000 and 2000 AU, a good scale
for finding low-mass multiple systems \citep{survey}.
Additionally, the interferometric data resolve out the very large-scale
emission from the molecular core, allowing us to probe the localized
density enhancements \citep{looney2003}.
Along with SCUBA dust continuum emission and infrared images, we
can explore star formation toward \bdstar by analyzing the
dense molecular gas, dust, and stellar distribution.

%In what follows, the morphology, kinematics, and
%column densities and masses derived from the CS emission are presented;
%and estimates of the total column density and obscuration towards
%this region are derived from the continuum emission at 98 GHz.

\subsection{Dense Gas and Dust Morphology}

\begin{figure}
\begin{center}
\includegraphics{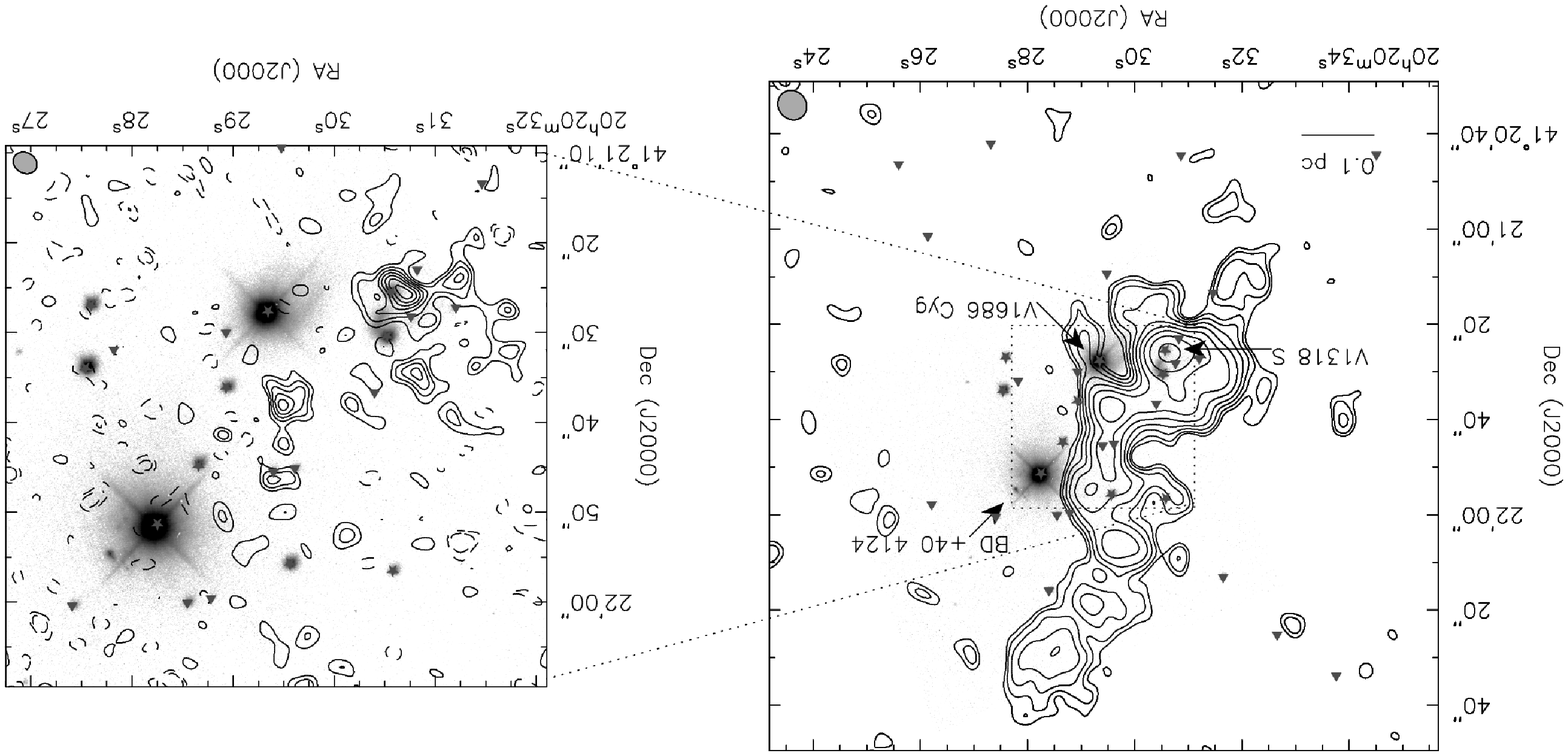}
\vspace{10cm}
\caption{
CS (2-1) emission toward the \bdstar young cluster
overlaid on a
smaller adaptive optic H-Band image \citep{ric}.  
Both CS maps are made from the same data
with different weighting.  On the left is the low resolution
image emphasizing the
large-scale structures.  The noise is 0.2 Jy/beam km/s.  The contours
are logarithmically spaced (increment $\sqrt{2}$) from 2 to 64 times the noise.
The negative contours, presumably from resolved out large-scale emission,
are not shown to simplify the image.  The beam, shown in the lower-right
corner, is 6$\farcs$42 $\times$ 5$\farcs$83 with a PA of 40\arcdeg.
The dotted box indicates the zoomed field of the high resolution map
(right).  
%Note that the H-Band field is much smaller than the CS field.
The high resolution image emphasizes the smaller-scale clumps of
material traced by CS.  The noise is 0.29 Jy/beam km/s.  The contours
are linearly spaced from 2 to 7 (positive and negative) times the noise.
The beam, shown in the lower-right corner, is 2$\farcs$76 $\times$
2$\farcs$27 with a PA of 62$\arcdeg$.  The symbols indicate IR sources:
the star symbols indicate IR sources with optical counterparts, and
the triangle symbols indicate IR sources without an optical counterpart
(presumably more embedded objects, thus younger).
}
\label{cslow}
\end{center}
\end{figure}

Figure \ref{cslow} illustrates the velocity-integrated maps
of the CS~(2-1) emission overlaid on an H-band adaptive optics observation
of the core stellar components \citep{ric}.
One of the striking morphological features of the CS emission
is the significant molecular gas
along a ridge running nearly north-south with an extent of $\sim$0.4 pc.
This ridge of molecular
emission is clearly aligned with the large amount of
dust continuum emission detected in the JCMT
SCUBA observations at $\lambda$~=~850~$\mu$m (Figure \ref{scuba}).
The two main peaks in the SCUBA image corresponds to the peaks
labeled A and J in the CS maps.
A comparison of the two maps shows that the CS emission is indeed
tracing the densest regions, or the molecular cloud core.
% where star formation
%may be ongoing.  
There are arguably 16 distinct clumps that can be recognized
in the velocity integrated maps (Figures \ref{cslow} \& \ref{scuba}).
% that may
%harbor protostellar systems.

Figure \ref{cont} shows the corresponding dust continuum emission at
$\lambda~=~3.1$~mm for the two resolutions.
As for the case of the $\lambda~=~850~\mu$m continuum, the
two main peaks coincide with clumps
A and J in the CS maps.
At high-resolution, the bright peaks of the CS and continuum
emission are centered near the close binary pair V1318, suggesting
that this region is where the youngest members of the cluster
are located.
The two brightest stars in the H-band observation of Figure \ref{cslow}
(\bdstar and V1686, spectral types B2 and B5, respectively) do not have any
dense gas or dust associated with them: both are only at the edge of the CS
emission. This fits in the picture that massive stars reveal
themselves quickly.
However, it is difficult to explain the lack of an \hii region
near \bdstar if the star is ablating the edge of the dense gas.

%The masses of the clumps range from 0.5 to 2.0 solar masses, similar
%to observed envelope masses in low-mass Class 0 protostars.
%This suggests that low-mass star formation may still be ongoing,
%producing more low-mass
%companions to the B-stars.

\subsection{Dust Structures}

In Figure \ref{cont}, the $\lambda$~=~3.1~mm continuum
emission \citep[also detected by][at lower resolution]{james97}
is overlaid on the H-Band continuum.  The low resolution image 
shows that many of the continuum peaks are cospatial with the CS
peak emission.  
Under the assumption of optically thin, isothermal dust emission, the
mass can be calculated using
F$_{\nu} = B_{\nu}(T_{dust})\kappa_{\nu}M/D^{2}$.
$B_{\nu}(T)$ is the Planck function, $T_{dust}$ is the temperature of the dust,
$\kappa_{\nu}$ is the dust mass opacity, M is the mass of gas and dust, and
D is the distance to the source.
Note that the derived mass scales roughly linear with 
the assumed dust temperature; we use $T_{dust}$ = 25 K.
For the dust mass opacity
\citep[e.g.,][]{draine90,pollack94}, we adopt a $\kappa_{\nu}$ which
is consistent with other works \citep[e.g.,][]{beckwith91,survey}:
$\kappa_{\nu}$ = 0.1($\nu$/1200 GHz) cm$^{2}$ g$^{-1}$, corresponding to
$\kappa_{\nu}$ = 0.008 cm$^{2}$ g$^{-1}$ and 
$\kappa_{\nu}$ = 0.029 cm$^{2}$ g$^{-1}$ in the BIMA and
SCUBA bands, respectively.
Although we
do not expect this simple model to give accurate masses, it provides
rough estimates that are adequate for qualitative comparisons and are
arguably within a factor of 2 of the likely total mass.

The two brightest continuum peaks in the low resolution image of
Figure \ref{cont} are 48.2 $\pm$ 3.7 mJy and 17.6 $\pm$ 3.4 mJy.
Using the above formalism, their 
estimated masses are 4.2 $M_\odot$ and 1.5 $M_\odot$, respectively.
In the high resolution image the bright continuum emission 
peak is resolved and
well fit as an extended Gaussian source
of 3100 by 1500 AU, assuming a distance of 1~kpc,
with a measured flux density of
38.6 $\pm$ 4.3 mJy, or a derived mass of 3.4 $M_\odot$.
The extinction (A$_V$) at the center of the high resolution
peak was derived 
%We can also derive an extinction (A$_V$) at the center of the high resolution
%peak 
by converting the peak brightness of 22.9 $\pm$ 1.8 mJy/beam
into an H$_2$ column density.
% using the same assumptions.
The mass in the peak beam is 2 M$_\odot$/beam.
Using a hydrogen mass abundance ratio (X$_H$) of 0.7,
that mass is equivalent to a column density of 
N$_{H_2}$~=~5.25$\times~10^{23}$~cm$^{-2}$ in the beam, again using
a distance of 1~kpc.
%, assuming
%that all the hydrogen is molecular in the core.
Assuming that R$_V$ = 3.2, we can use
the relationship A$_V$~/~N(H$_2$)~=~$1.33~\times~10^{-21}$~cm$^2$~mag, 
after \cite{bohlin1978}.  The extinction of the compact structure
at the peak is A$_V$ $\sim$ 700 mag, a very dense region indeed.

Finally, the two brightest peaks in the SCUBA map (Figure \ref{scuba}) 
are 2.83 $\pm$ 0.045 Jy/Beam and 1.26 $\pm$ 0.045 Jy/Beam. 
%Again, the uncertainties listed are the statistical uncertainties;
%the estimated absolute flux calibration is 10\%.
%Using the above formalism,
An estimated point mass of 6.9 $M_\odot$ and 3.1 $M_\odot$,
respectively, is determined.  
The total flux from the whole region is 28.2 $\pm$ 0.3
Jy, or 68.5 $M_\odot$, a factor of 4 less than the estimated
molecular mass of the cloud from single-dish measurements.
However, this mass estimate assumes a temperature of 25 K,
which is probably an overestimate for the average
temperature of the whole cloud, underestimating the total mass.  
In fact, a temperature estimate of 10 K for the whole
region would match the mass of material detected in
single dish observations-- $\sim$300~$M_\odot$.

\subsection{Gas Kinematics}
 
\begin{figure}
\includegraphics{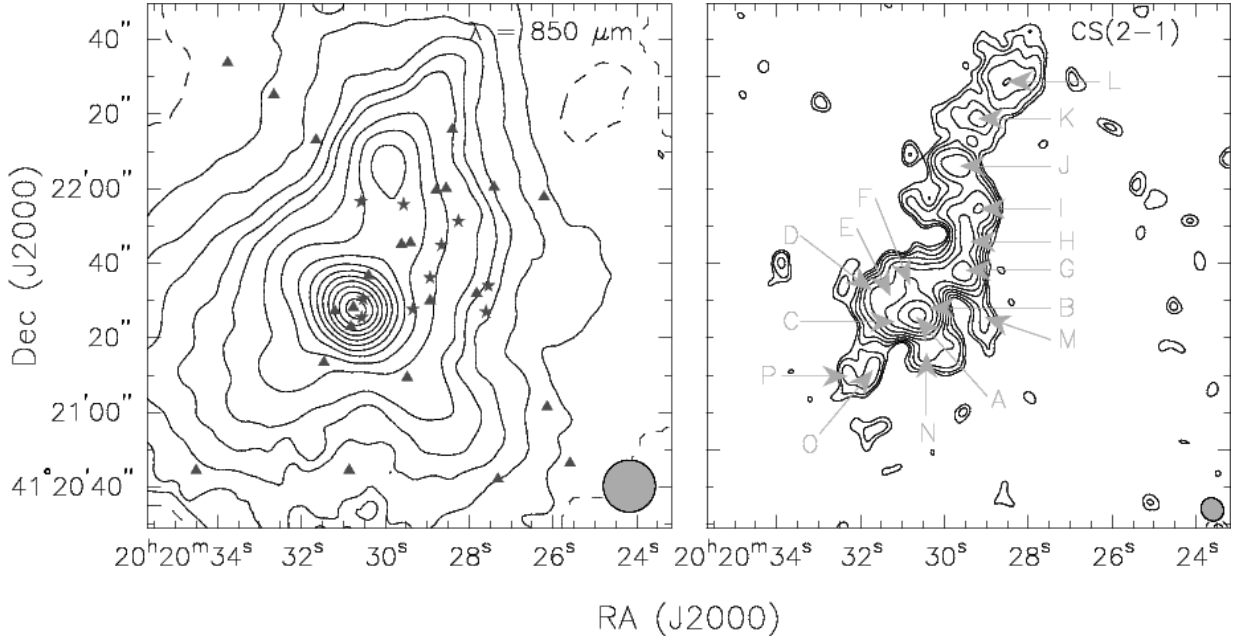}
\vspace{4cm}
\caption{
$\lambda$~=~850~$\mu$m continuum emission toward
the \bdstar young cluster in the left panel is
archival SCUBA observations.
The contours are in steps of 2 from 2 to 10 and in steps of 5 from
10 to 60 times the noise of 0.045 Jy/beam.
The beam shown at the lower right is 14.7$\arcsec$.
The symbols are the same as in Figure \ref{cslow}.
The right panel is the low resolution CS emission from Figure \ref{cslow}
with the 16 clumps labeled.
}
\label{scuba}
\end{figure}

Sixteen locations (see Figure \ref{scuba})
were chosen based on the clumpiness of the CS emission
in the low and high resolution integrated intensity maps.
Figures \ref{spec_low} \& \ref{spec_hi} show the CS line emission
integrated over a beam element at the 16 locations 
in the low and high resolution maps (see Figure \ref{cslow}), respectively.
In general, the upper spectra are from the core region (containing the
V1318 system) where the most active star formation is occurring, the
middle spectra are from the northern spur, and the bottom spectra
are from the fringe regions.

Table \ref{fits} lists the single Gaussian fits to the low spatial
resolution spectra as labeled in Figure \ref{scuba}.
There is a noticeable difference in the spectra when comparing the
core region and the northern spur (also see Figure \ref{lines}).
%Clearly, the star formation activity in the central core
%is influencing the dense gas nearby.
%The difference in linewidth is clear indication of the effect of active
%star formation on the molecular cloud of a cluster.  
First of all, the linewidths
decrease from a maximum value of 2.78 km/s near the CS peak to 1.07 km/s
at the northern clump K.  In general, the linewidth
generally decreases as the distance from the central core.
Secondly, there are two clear linear correlations 
in the fitted Gaussian center velocity of the line.
In the core (approximately -40\arcsec to 15\arcsec), 
the velocities increase from 7.2 km/s to 8.3 km/s, and in the
northern spur (approximately 0\arcsec to 55\arcsec), again
the velocities generally increase from 6.7 km/s to 7.5 km/s.

%The core condensations (peaks A through F) are generally 
%centered at velocities
%of 7.3 to 8.2 km/s, similar to the cloud velocity as measured in
%CO \citep[$\sim$8 km/s,][]{loren77,canto84} and 
%$^{13}$CO \citep[7.7 km/s][]{canto84}.
%The northern spur condensations (peaks G through K) are
%systematically centered at velocities 6.7 to 7.3 km/s.

What is causing the two systematic velocity trends?
One possibility is an overall rotation of the initial molecular
cloud that formed the cluster, but there are two separate
linear correlations in Figure \ref{lines}.
Another possibility is that the emission from the Northern component 
velocities are contaminated by outflows.  However,
at the highest resolution, where the line
profiles are less influenced by outflow emission, 
the peak velocities are not significantly different
(Figure \ref{spec_hi}).
Finally, the most compelling argument is that
the velocity gradients
are real and that the elongated cloud consists of two components.

Indeed, the kinematics clearly suggest that the molecular
material surrounding the cluster consists of two
clouds and that the systematic velocity gradients are due to rotation.
The clumps P, O, N, C, A, B belong to cloud
1 and  the clumps G, H, I, J, K, L to cloud 2.  
Assuming a rigid
rotation cloud model, linear fits of the velocities
over the offsets yield gradients of 0.033 and 0.015~km/s/$\arcsec$ 
for cloud
1 and cloud 2, respectively. The linear fit describes the data well
for both clouds, as the scatter (RMS) of the residual are about
0.1~km/s compared to the initial scatter of 0.38 and 0.32~km/s, 
respectively. The clumps E and F are associated to cloud 1 as their
velocities match the other clumps, but are excluded from the fit
as they are spatially separated from the rest of cloud 1 and
projection effects may place them at unsuitable offsets.

%The clumps P, O, N, C, E, A, F, and B belong to cloud 1,
%and the clumps G, H, I, J, K, L belong to cloud 2.
%A linear fit of the velocities for the two clouds yields
%0.03 km/s/$\arcsec$ and 0.01 km/s/$\arcsec$, respectively.
%The residuals from this analysis produced, using
%both clouds, an RMS of 0.18 km/s, compared to 
%0.42 km/s and 0.29 km/s for cloud 1 and cloud 2, respectively.
%Of course, the residuals of cloud 1 could be further reduced by not including 
%clumps E and F, but there is no clear reason to neglect those clumps.
%A similar analysis of the FWHM of the lines gives
%0.028 km/s/$\arcsec$ and -0.003 km/s/$\arcsec$ for cloud 1 and
%cloud 2, respectively. 
%However, the residuals from the fit do not greatly reduce the RMS,
%0.38 km/s to 0.28 km/s and 0.21 km/s to 0.20 km/s for cloud 1 and
%cloud 2, respectively.

The evidence of two kinematically separate clouds
implies that the northern spur is the remnant of a 
cloud that collided with the southern condensation. 
This is a very
intriguing result, as these two clouds may be the
remnant signature of the, so far, unknown trigger of star
formation and account for its high star-formation efficiency.
This may be compared to the
recent ideas on the creation of molecular clouds from colliding
streams of atomic hydrogen \citep[e.g.,][]{vazquez2003,heitsch2005}, but the
material (CS compared to hydrogen), size-scales, and
probable velocity linewidths are very different. 

In addition to the collision of two clouds, 
there is a clear distinction in Figure \ref{lines}
around the immediate region of clump A-- the 
most active region of star formation in the cluster.
The increased linewidths of the region, coupled
with possible fluctuations in the peak velocities
(i.e. clumps E and F),
are arguably the effect of star formation in the V1318 Cyg
region; an intermediate star churning up the intragroup environment
with winds and outflows (see the following section).
On the other hand, the dense gas nearest BD+40\arcdeg 4124 and V1686
Cyg displays no evidence of disruption, indicating that the disturbed
material is processed quickly into the star, 
removed from the system, or relaxes
fairly quickly.  

There are two spectra that are not well fit by the Gaussian
models in Figure \ref{spec_low} -- clumps D and M.
Because of this, they were not included in the above discussion. 
In the case of D, it is unclear if the dip in the center is
from self-absorption or from two separate clumps.
However, due to the unusual velocities for the two clumps and the
reduction in brightness compared to the other central regions,
we prefer the self-absorption scenario.
There is a slight blue asymmetry to the emission
that may be an indication of an infall signature from 
an embedded object \citep[e.g.][]{walker94}.
However, at high resolution (Figure \ref{spec_hi})
only the blue component is observed.
Also, as seen in Figure \ref{scuba}, there is an extension in the submillimeter
continuum toward clump D that has no stellar sources, perhaps
indicating a region of dense material.
For clump M, the spectra clearly show the classical
signature of a P-Cygni profile.  
Clump M may be affected by the Herbig Ae/Be source V1686 Cyg.
If this interpretation is correct, the line is absorbing against the
large-scale continuum; there is no detected compact emission
by the interferometer, but the large continuum source (e.g. Figure \ref{scuba})
is resolved out.

\subsection{Outflow Feature}

In Figure \ref{spec_low}, one can distinguish components from an outflow
at 4 km/s and 11 km/s.
Figure \ref{outflow} shows the velocity integrated emission over
the blue (2.462 - 4.854 km/s) and
red components (10.236 - 12.03 km/s) of the clump A wings.
This is consistent with the emission being the low-velocity
dense component of the detected large-scale outflow
detected in CO \citep{palla95}.
It is also interesting to note that the source of the outflow
is better attributed to the maser object than the optical/IR 
southern binary (Figure \ref{cont}, high resolution).

\subsection{Column Densities and Masses of the Clumps}

From the observation of a single transition, the
total number density, $n_{H_2}$, and kinetic temperature, $T_K$, of the
emitting region can not be derived. 
However, the {\it range\/} of
molecular column densities that can account for the observed
brightness temperatures can be estimated. By computing a grid of
Large-Velocity-Gradient (LVG) models with a range of $n_{H_2}$, $T_K$, and
molecular column densities $N_{CS}$ as input, and using this grid to fit
the observed brightness temperatures, a range of $N_{CS}$ is found
that can account for the emission. The usefulness of the obtained
molecular column densities depends on the range limits.
Note that, because only one transition is at hand, only one of $n_{H_2}$, $T_K$,
or $N_{CS}$ can be deduced in this manner. $N_{CS}$ is chosen, as 
the amount of molecular gas is particularly interesting and
%to deduce $N_{CS}$ from $n_{H_2}$ one would
%have to assume a molecular abundance and a column length-- assumptions
%that would introduce further uncertainties. Also, it turns out that
the range in $N_{CS}$, consistent with the observed emission, is always
narrower than the corresponding ranges in $T_K$ and $n_{H_2}$.

The output range of $N_{CS}$ can be narrowed further by considering only
those models for which the optical depth of the transition under
consideration is $\la 2$. This constraint is based on the fact that
for $\tau \ga 2$ the validity of the LVG approximation is
uncertain-- at best. In the case at hand, the observations
of \cite{fuente90} provide an additional constraint. They deduced
$T_K$ from the (1,1) and (2,2) lines of $NH_3$, and found values
ranging from 18 K to 27 K.
Therefore, the input values of $T_K$ can be 
chosen to be in  the range 15--30 K.

A grid of LVG models was computed with 9 kinetic temperatures in the
range 10--100 K, 9 values of $n_{H_2}$ in the range
$10^4$--$10^8$ cm$^{-3}$, and $20$ CS column densities in the range
$10^{13}$--$10^{15}$ cm$^{-3}$-- logarithmically spaced in all three
variables. 
The obtained values of $N_{CS}$ (for $15 K \le T_K \le 30 K$ and
$\tau \le 2$;  and using the values of
$\Delta v_{FWHM}$ and their uncertainties in Table \ref{fits} are
presented in Table \ref{lvg}.  

With this range of $N_{CS}$ and assuming $X_{CS} = 10^{-9}$ 
for the CS to H$_2$ abundance,
the mass range of the clumps (Table \ref{lvg}) were deduced from 
\begin{equation}
  M = \mu m_{H}\, \frac{N_{CS}}{X_{CS}}\, (1.13\, \theta_a \theta_b D^2)
\end{equation}
where $\mu=2.33$ is the mean molecular weight, $m_{H}$
is the mass of the hydrogen atom, and the last term is
the area subtended by the beam at a distance D ($1\,{\rm kpc}$).
This can be compared to the virial mass as calculated from 
\begin{equation} 
   M_{\rm vir} = \frac{5}{3}\,
   \left(\frac{3}{8\ln 2}\right)^{1/2}\,
    \frac{\Delta v_{FWHM}^2}{G}\, R,
\end{equation}
where R is the physical radius subtended by the beam.

Perusal of Table \ref{lvg} shows that, for each emission peak, the range of
$N_{CS}$ is fairly narrow and the typical column density of
H$_2$ is $\la 10^{23}$ cm$^{-2}$, consistent with our
estimate from the continuum data. 
It is not surprising that the column densities are an order of
magnitude larger than that derived from single dish
$^{13}$CO data \citep{canto84} and a factor of two larger than the
single dish value estimated using $^{12}$CO and $^{13}$CO
\citep{lynne95}.
The higher spatial resolution of interferometric observations,
6$\arcsec$ compared to 66$\arcsec$ and 15$\arcsec$ for \cite{canto84}
and \cite{lynne95}, respectively, allow the distinction of
high density clumps and not averaging them out.
However, it is important to note that the molecular hydrogen column
densities presented on Table \ref{lvg} were derived assuming a CS to H$_2$
abundance of $10^{-9}$, which could be too large by a factor $\sim 5$.

%Indeed, the estimated masses of the clumps show that the amount of mass
%is sufficient to form stars, although the 
Finally, comparison of the masses derived from the LVG analysis and
those computed with the virial theorem show that in all cases
$M<M_{\rm vir}$. Given the crudeness of the LVG analysis presented
here, it is difficult to draw any definitive conclusion from this
result. 

\subsection{Stellar Population}

Figures \ref{cslow}, \ref{scuba}, and \ref{cont}
also show the distribution of the known stellar
population \citep[][although the observations were not deep: 100\% complete
at K=15.3]{lynne95}
with respect to the dust and dense gas.
The star symbols are IR sources with optical counterparts
while the triangles are IR sources without any known optical counterparts.
IR excesses are found in 
100\% of the known stellar population
\citep{lynne95}; both the low-mass and the high-mass stars are young.
This suggests that low-mass and high-mass star formation occurred
quickly-- $<$~1 Myrs with high formation efficiency.

However, the evolutionary stages of the sources are difficult
to estimate until more is known about
their spectral energy distributions and the foreground extinction.
Nonetheless, one can compare the extinction of the sources estimated from
their position on a color-magnitude diagram \citep[e.g. Figure 8 in][]{lynne95}
to the extinction estimated from the CS line.
The estimated extinction in the clumps range from
A$_V$ = 14 to $>$~100 mag on the molecular ridge of material.
For the optically visible IR sources in Figure \ref{cslow},
the extinctions are typically around A$_V \sim$ 1
regardless of location-- on-ridge or off-ridge.
The major exception is V1686 Cyg, 
which has a high extinction estimate \citep{lynne95} and is near
the molecular ridge.
Similarly, the majority of the  
IR sources without optical counterparts have 
extinctions that are typically around A$_V \sim$ 13.
As the extinction is independent of the source field position,
the effect is probably due to local conditions at the source, not
its location within the dense molecular gas.
In general, this evidence leads to the assumption that typically,
the IR only sources are slightly embedded protostars (i.e. Class I/II)
while the optically visible sources are 
young stars similar to T Tauri systems (i.e. Class II/III).  

In that case, 19 out of a total of 31
stars in our field (7/10 of the Class II/III objects and 12/21 of the
Class I/II objects) are closely associated with dense gas.
It can be argued that the majority of the protostars are very closely
associated with the CS ridge and that the Class II/III sources
are forming closer to the more massive stars, neglecting projection effects.
If star formation started approximately coevally, this
suggests that the massive stars hasten the evolution of the
objects nearest them.
In other words, the massive stars reveal the low-mass stars from their
dust cocoons, making them appear more evolved.
On the other hand, as the ridge of intracluster material is clumpy on the size-scale
expected for protostellar cores, many of the numerous dense gas clumps
in Figure \ref{cslow} may be harboring deeply embedded stellar counterparts.
For example, a low-mass Class 0 system would not be seen
in the IR as the dense envelope extincts light at wavelengths less than
10 $\mu$m.
Indeed, in this paper we will show evidence for an intermediate-mass
deeply embedded protostar; future FIR observations will help
identify other such embedded objects.
Star formation is not completely coeval in this region.

Finally, it is clear that the IMF
of this cluster is unusually top-heavy based on the known
star counts \citep{lynne95}.
Does the IMF vary locally with
an integrated value over clusters?  Or is the hidden population
of low-mass stars enough to compensate.
A census of the hidden low-mass stellar population would be
extremely helpful.
Indeed, this may be addressed with Spitzer observations of this region.

\section{A New, Deeply Embedded Protostar}

\begin{figure}
\includegraphics{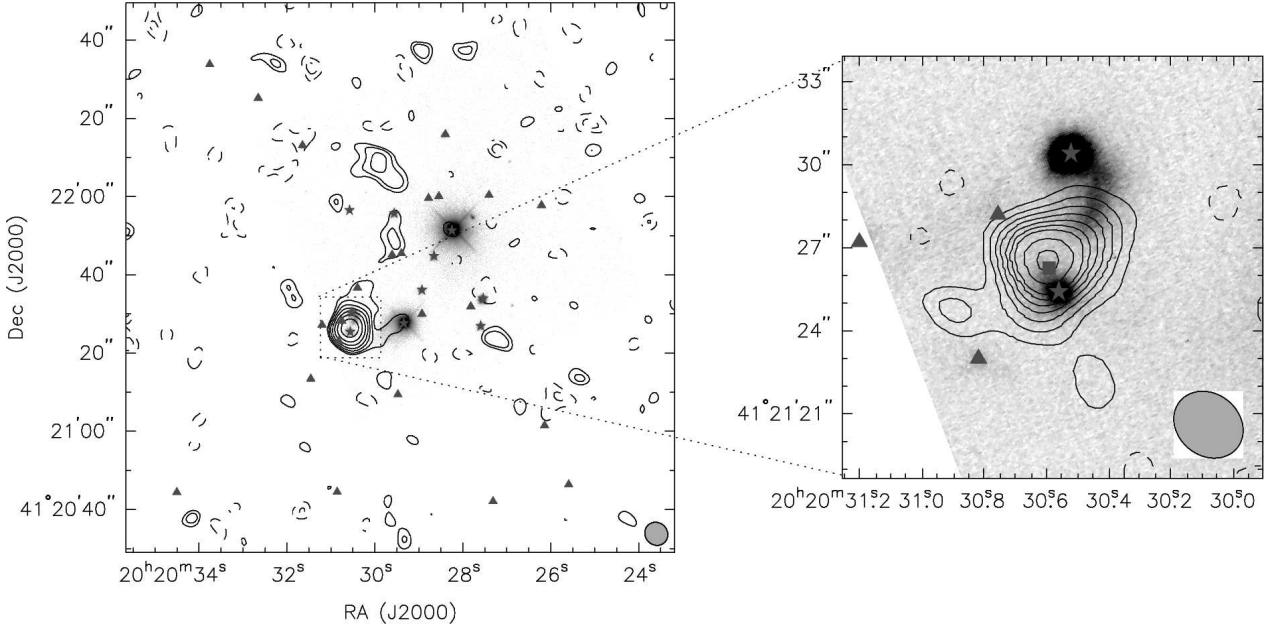}
\vspace{8cm}
\caption{
$\lambda$~=~3.1~mm continuum emission toward 
the \bdstar young cluster
overlaid on a smaller adaptive optic H-Band image \citep{ric}.
Both maps are made from the same data
with different weighting.  On the left is the low resolution
image emphasizing the
large-scale structures.  The noise is 1.6 mJy/beam.  The contours
are logarithmically spaced (increment $\sqrt{2}$) from 2 to 32 times the noise;
negative values are indicated by dashed contours.
The beam, shown in the lower-right
corner, is 6$\farcs$12 $\times$ 5$\farcs$68 with a PA of 45$\arcdeg$.
The dotted box indicates the zoomed field of the high resolution map
(right).
The high resolution image emphasizes the compact emission.
The noise is 1.8 mJy/beam.  The contours
are linearly spaced from 2 to 20 (positive and negative in steps
of two) times the noise.
The beam, shown in the lower-right corner, is 2$\farcs$67 $\times$
2$\farcs$24 with a PA of 52$\arcdeg$. 
The triangle and star symbols are the same as in Figure \ref{cslow}, but
the solid square indicates the position of the VLBA water maser source.
Inset is a zoom of the V1318 region using a different colormap
on the H-band data to emphasize the diffuse structure.
Note that the H Band spur off the northern component is a real
scattered light feature.
}
\label{cont}
\end{figure}

\begin{figure}
\includegraphics{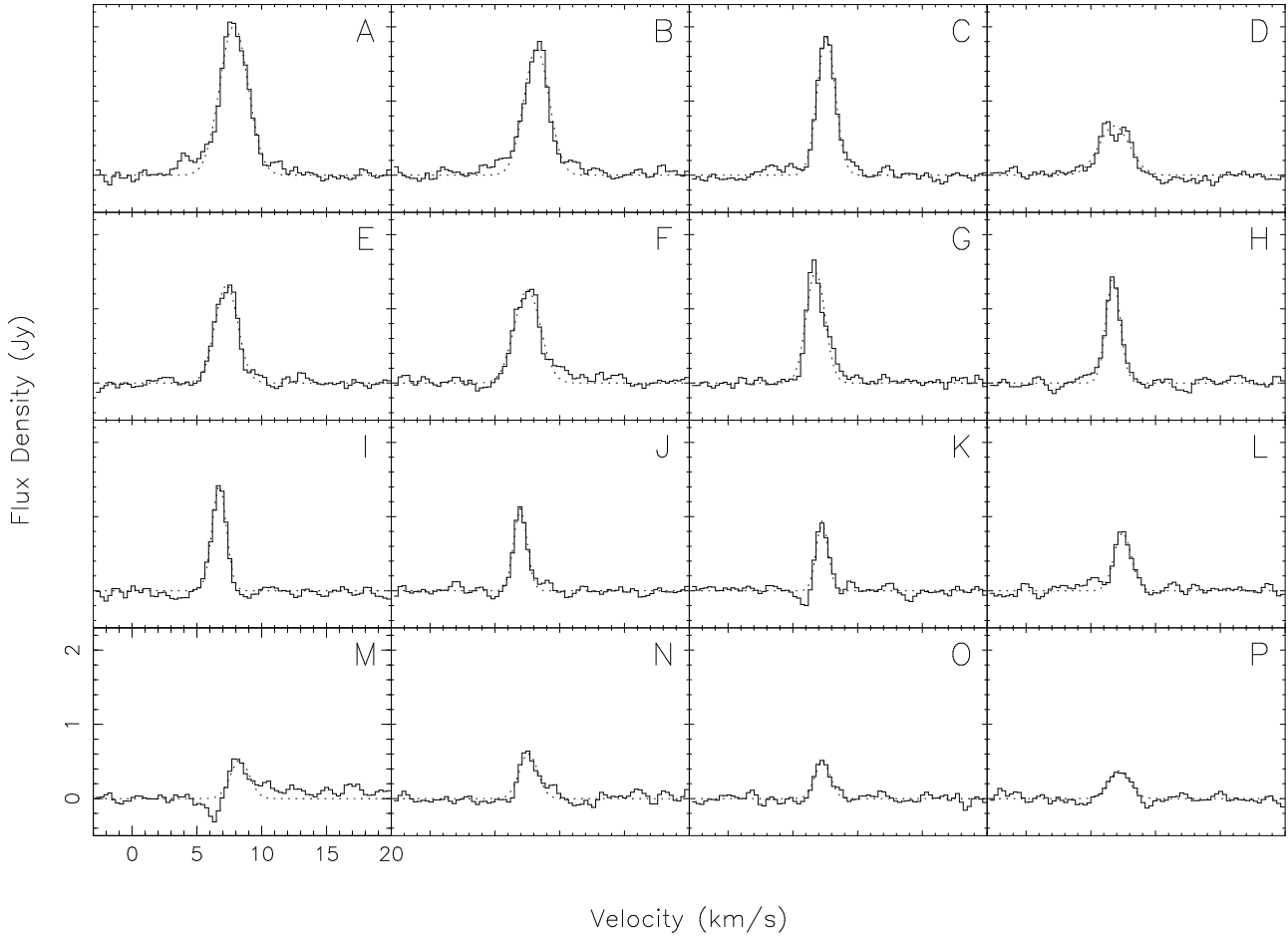}
\vspace{8cm}
\caption{
Beam averaged spectra of the 16 clumps in the low resolution image.
The dashed line is the best fit Gaussian in each case, as labeled
in Table \ref{fits}.
}
\label{spec_low}
\end{figure}

\begin{figure}
\includegraphics{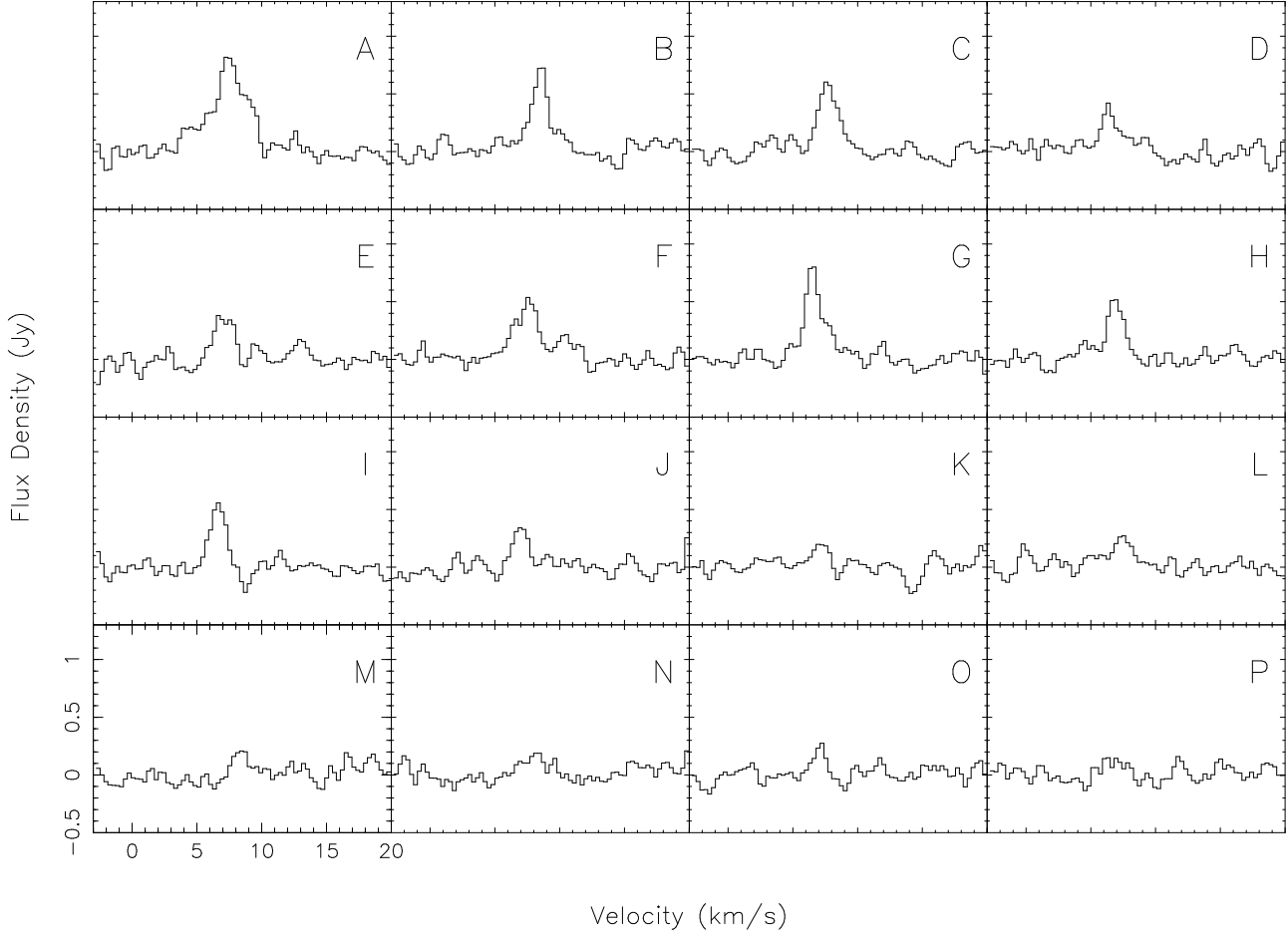}
\vspace{8cm}
\caption{
Beam averaged spectra of the 16 clumps in the high resolution image.
}
\label{spec_hi}
\end{figure}

\begin{figure}
\includegraphics{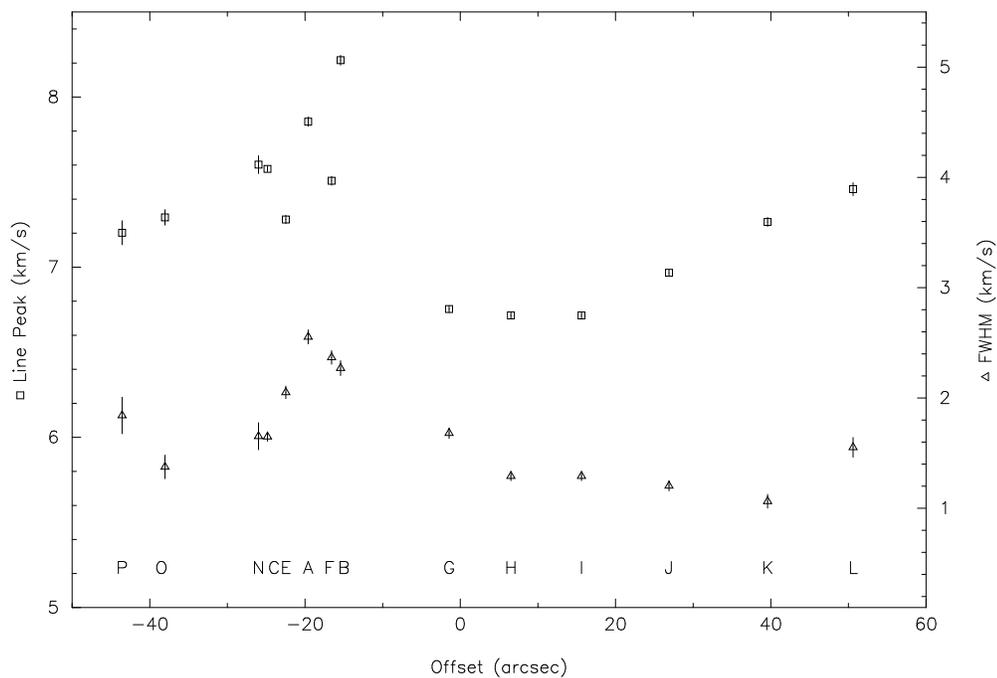}
\vspace{3cm}
\caption{
Comparison of the low resolution spectra peak velocity and linewidth
with offset from the pointing center (near peak G) in arcseconds.
Clumps D and M are not plotted as their Gaussian fits are not acceptable.  
The square symbols are the line center velocities of the remaining 
14 best fit Gaussians, as indicated by the left-hand axis.
The triangle symbols are the corresponding FWHM of the
best fit Gaussians, as indicated by the right-hand axis.
The error bars are the Gaussian fit error on the parameters.
This assumes that the spectra are well defined as Gaussians.
At the bottom, each clump is labeled with the associated letter.
}
\label{lines}
\end{figure}

\begin{figure}
\begin{center}
\includegraphics{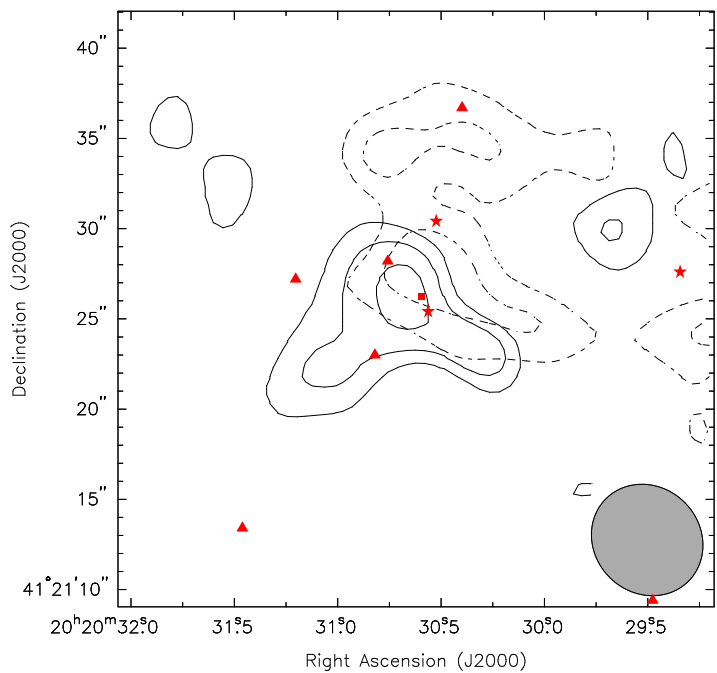}
\vspace{8cm}
\caption{
CS (2-1) emission toward the \bdstar young cluster 
in the excess
velocities of clump A (Figure \ref{spec_low}).
The velocity range of 2.462 - 4.854 km/s is the blue component
(solid contours) and the velocity range of 10.236 - 12.03 km/s
is the red component (dashed contours).
The noise is 0.12 Jy/beam km/s.  The contours
are 2, 3, and 4 times the noise.
The beam, shown in the lower-right
corner, is 6$\farcs$42 $\times$ 5$\farcs$83 with a PA of 40$\arcdeg$.
The triangle and star symbols are the same as in Figure \ref{cslow}, but
the solid square indicates the position of the VLBA water maser source.
%This is consistent with this emission being the low-velocity
%dense component of the detected large-scale outflow
%detected in CO \citep{palla95}.
}
\label{outflow}
\end{center}
\end{figure}

The resolved dust structure in Figure \ref{cont} could be a flattened
circumstellar disk, but we suggest that, due to its size, this structure
is more likely a flattened envelope that surrounds a protostar; in
particular, an unknown deeply embedded intermediate mass protostar.
As seen in the zoomed inlay of Figure \ref{cont},
the peak of the dust emission is offset by 1$\farcs$1
to the northeast from the southern member of the V1318 binary.
This offset is inconsistent with the observational
uncertainties (expect $\sim$ 0$\farcs$5).
In addition, this location is precisely where
H$_2$ emission is detected \citep{ric}
and is coincident within $\sim$ 0$\farcs$25 of
the recent water maser emission observations
\citep[the square symbol in Figure \ref{cont},][]{marvel06}.
The high resolution ($\sim$ 2$\arcsec$) CS emission
shown in Figure \ref{cslow} also has an offset similar to the continuum.
In fact, although it can not be ruled out by these data alone, if
the continuum and its associated extinction (A$_V$ $\sim$ 700)
is physically associated with the southern source, i.e.
the extinction is due to the massive envelope of the embedded objects,
we must have a unique geometric observing position to detect the southern
binary object in the optical at all.
Rather, due to the discussion above,
we suggest that the compact continuum emission and high resolution CS
emission are actually tracing a younger, more embedded object,
probably an intermediate-mass protostar.
Indeed, a young cluster population appears to be lurking inside the dense gas.
The southern component of the binary, must be a star at
an older stage of evolution at an unfortunate projection.

\section{Conclusions}

In this paper, we have probed the dense gas and dust in 
the small cluster associated with the intermediate-mass
young star \bdstar.
The dense gas, as traced by CS J = 2 $\rightarrow$ 1, 
consists of $\sim$16 clumps that have
the morphology of a dense molecular ridge 
about $\sim$ 0.4 pc in size.
By a simple LVG analysis of the clumps, we find that the
typical estimated mass of the clumps is $\sim$ 0.5 M$_\odot$.
Although it is impossible with this data 
to predict which, if any, of these clumps may contain
a young, deeply embedded population of protostars, it seems likely
that some clumps do.

In particular, we resolve a flattened dust structure,
3100 $\times$ 1500 AU in size that is 1.1$\arcsec$
offset from the southern component of the
V1318 Cyg binary system, but well aligned with
the water maser position.
By modeling the circumstellar emission as
arising from optically thin isothermal dust, a
mass of 3.4 $M_\odot$ is estimated for the structure.
The V1318 Cyg binary system is
optically visible; however, the extinction is 
expected to be around A$_V$ $\sim$ 700 mag. 
With so high an extinction, the central source should not be
visible.  If the optically visible V1318 S is the source
of the continuum emission, it would require an unusual geometry
that is not evident in the continuum emission.
Based on the offset and the visibility of V1318 Cyg-S,
we suggest that this object is a hereto unknown
young embedded object and not V1318 Cyg-S.
The structure, based on its mass, is probably a flattened
circumstellar envelope which contains 
an intermediate mass, deeply embedded protostar.
Indeed, star formation is still ongoing in this cluster.

However, we posit that, in general, star formation of this small cluster was 
coevally triggered
by the collision of two clouds.  The dense gas detected in this paper
is essentially the remnants of the two original cores.  The main
reason for this conclusion is that the systematic velocity of the
CS~($2\rightarrow 1$) transition suggests the presence of 
two independently rotating clouds, distributed spatially: 
cloud 1 is the center of the southern
star forming activity and cloud 2 is the northern spur.  Currently,
cloud 1 is the center of the dense dust and gas in the cluster.
Triggered star formation would probably 
explain the high star formation efficiency
in the system.  
%In addition, the \bdstar cluster
%could be an isolated example of colliding
%streams \citep[e.g.,][]{vazquez2003,heitsch2005}.

Finally, 
there is evidence, e.g. the linewidths of the V1318 Cyg region,
that the young intermediate mass stars in the core are
disrupting the dense gas, which interferes with 
coeval lower-mass star formation.
However, it must also be noted that this argument also implies that the
previous stages of star formation, in particular the formation
of \bdstar and V1686 Cyg, did not disrupt the kinematics
of the residual clouds significantly.
In other words, the gas that was affected by their formation
has been processed into the star, removed from the system, or
relaxed fairly quickly.
Another example of this is that the more evolved low-mass
protostars are more likely to be near the more massive young stars.
If the system was triggered by colliding clouds, then
the majority of the low-mass stars are of similar ages.
That implies that the presences of the higher mass
protostars either hastens the lower-mass evolution, or that the
more massive stars oblate the circumstellar
structures of the lower-mass siblings, making them appear older.

\acknowledgements{
We thank Lanie Dickel for careful discussions,
Jason Kirk for assistant with the SCUBA data, and especially
Lynne Hillenbrand for her help with this work.
L.W.L., S.W., and M.H. acknowledge support from the Laboratory for
Astronomical Imaging at the University of Illinois and NSF AST 0228953.
The JCMT Archive project is a collaboration between the Canadian Astronomy
Data Centre (CADC), Victoria and the James Clerk Maxwell Telescope
(JCMT), Hilo. Funding for the CADC's JCMT Archive effort is provided by
the National Research Council of Canada's (NRC) to the Herzberg Institute
of Astrophysics.}

%\bibliographystyle{apj}
%\bibliography{bd40}

\input{ms.bbl}
\clearpage
\clearpage

\clearpage

\clearpage

\clearpage

\clearpage

\clearpage

\clearpage

\input{tab1.tex}
\input{tab2.tex}

\end{document}

%% file: tab1.tex
\begin{deluxetable}{lrccc}
\tablewidth{0pt}
\tablecaption{Fit Parameters}
\tablehead{
  \colhead{Label} & \colhead{Position} & \colhead{Peak} & 
  \colhead{velocity} & 
  \colhead{FWHM}\\
  \colhead{} & \colhead{$\arcsec$} & \colhead{Jy} & 
    \colhead{km/s} & \colhead{km/s}}
\startdata\label{fits}
A & -19.6 &2.03 $\pm$ 0.04 & 7.86 $\pm$ 0.03 & 2.55 $\pm$ 0.07\\
B & -15.4 &1.72 $\pm$ 0.04 & 8.22 $\pm$ 0.03 & 2.26 $\pm$ 0.07\\
C & -24.8 &1.87 $\pm$ 0.04 & 7.58 $\pm$ 0.02 & 1.65 $\pm$ 0.03\\
D & -26.8 &0.66 $\pm$ 0.03 & 6.81 $\pm$ 0.06 & 2.78 $\pm$ 0.15\\
E & -22.5 &1.33 $\pm$ 0.03 & 7.28 $\pm$ 0.02 & 2.05 $\pm$ 0.05\\
F & -16.6 &1.26 $\pm$ 0.05 & 7.51 $\pm$ 0.03 & 2.36 $\pm$ 0.07\\
G & -1.4  &1.52 $\pm$ 0.04 & 6.75 $\pm$ 0.02 & 1.66 $\pm$ 0.05\\
H &  6.5  &1.43 $\pm$ 0.04 & 6.72 $\pm$ 0.02 & 1.30 $\pm$ 0.07\\
I & 15.6  &1.43 $\pm$ 0.04 & 6.72 $\pm$ 0.02 & 1.30 $\pm$ 0.07\\
J & 26.9  &1.11 $\pm$ 0.04 & 6.97 $\pm$ 0.02 & 1.20 $\pm$ 0.05\\
K & 39.6  &0.94 $\pm$ 0.05 & 7.27 $\pm$ 0.03 & 1.07 $\pm$ 0.07\\
L & 50.6  &0.79 $\pm$ 0.04 & 7.46 $\pm$ 0.04 & 1.55 $\pm$ 0.08\\
M & -14.5 &0.52 $\pm$ 0.06 & 8.32 $\pm$ 0.10 & 1.71 $\pm$ 0.23\\
N & -26.0 &0.61 $\pm$ 0.04 & 7.60 $\pm$ 0.05 & 1.65 $\pm$ 0.12\\
O & -38.0 &0.51 $\pm$ 0.03 & 7.29 $\pm$ 0.03 & 1.38 $\pm$ 0.10\\
P & -43.6 &0.38 $\pm$ 0.03 & 7.20 $\pm$ 0.07 & 1.85 $\pm$ 0.17
\enddata
\end{deluxetable}
\clearpage

%% file: tab2.tex
\begin{deluxetable}{lcccc}
\tablewidth{0pt}
\tablecaption{Derived Column Densities and Masses}
\tablehead{
  \colhead{Position} & \colhead{$N_{CS}$} & \colhead{$N_{H_2}$} & 
  \colhead{M} & \colhead{$M_v$}\\
  \colhead{} & \colhead{log(cm$^{-2}$)} & \colhead{log(cm$^{-2}$)} &
	\colhead{$M_\odot$} & \colhead{$M_\odot$}}
\startdata\label{lvg}
A & 13.41-14.14 & 22.41-23.14 & 0.33-1.80 & 26.9\\
B & 13.35-14.09 & 22.35-23.09 & 0.29-1.58 & 21.2\\
C & 13.22-13.95 & 22.22-22.95 & 0.21-1.15 & 11.3\\
D\tablenotemark{\dagger} & 13.44-13.87 & 22.44-22.87 & 0.36-0.95 & 32.0\\
E & 13.31-13.52 & 22.31-22.52 & 0.26-0.42 & 17.4\\
F & 13.37-13.90 & 22.37-22.90 & 0.30-1.01 & 23.1\\
G & 13.22-13.85 & 22.22-22.85 & 0.21-0.90 & 11.4\\
H & 13.11-13.75 & 22.11-22.75 & 0.17-0.73 & 7.0\\
I & 13.11-13.75 & 22.11-22.75 & 0.17-0.73 & 7.0\\
J & 13.08-13.50 & 22.08-22.50 & 0.15-0.40 & 6.0\\
K & 13.03-13.45 & 22.03-22.45 & 0.14-0.40 & 4.7\\
L & 13.19-13.61 & 22.19-22.61 & 0.20-0.53 & 10.0\\
M\tablenotemark{\dagger} & 13.23-13.55 & 22.23-22.55 & 0.22-0.46 & 12.1\\
N & 13.22-13.64 & 22.22-22.64 & 0.21-0.55 & 11.3\\
O & 13.14-13.46 & 22.14-22.46 & 0.18-0.37 & 7.9\\
P & 13.27-13.48 & 22.27-22.48 & 0.24-0.39 & 14.2 
\enddata
\tablenotetext{\dagger}{Note that the derived
masses of D and M may not be valid if the J=2-1 CS line is
self-absorbed ($\tau >$ 2).}
\end{deluxetable}
\clearpage